\input{aipcheck}

\documentclass[
    ,final            
  ]
  {aipproc}

\layoutstyle{6x9}

\newcommand{\eqb}{\begin{equation}}  
\newcommand{\eqe}{\end{equation}}

\newcommand{\pperp}{p_{\perp}}
\newcommand{\pperpb}{p_{\perp 0}}

\newcommand{\ppar}{p_{\parallel}}
\newcommand{\pparb}{p_{\parallel 0}}
\newcommand{\gammaw}{\gamma_{*}}

\newcommand{\betaw}{\beta_{*}}
\newcommand{\betawb}{\beta_{*0}}
\newcommand{\pext}{p_{\rm ext}}
\newcommand{\av}[1]{\left\langle #1 \right\rangle}

\begin{document}

\title{High-energy emission from pulsar binaries}



\author{Iwona Mochol}{
  address={Max-Planck-Institut f\"ur Kernphysik, Saupfercheckweg 1, D-69117 Heidelberg, Germany}
}

\author{John G. Kirk}{
  address={Max-Planck-Institut f\"ur Kernphysik, Saupfercheckweg 1, D-69117 Heidelberg, Germany}
}

\begin{abstract}
Unpulsed, high-energy emission from pulsar binaries can be attributed to the
interaction of a pulsar wind with that of a companion star. At the shock between the
outflows, particles carried away from the pulsar magnetosphere are accelerated and
radiate both in synchrotron and inverse Compton processes. This emission constitutes
a significant fraction of the pulsar spin-down luminosity. It is not clear however,
how the highly magnetized pulsar wind could convert its mainly electromagnetic
energy into the particles with such high efficiency. Here we investigate a scenario
in which a pulsar striped wind converts into a strong electromagnetic wave before
reaching the shock. This mode can be thought of as a shock precursor that is able to
accelerate particles to ultrarelativistic energies at the expense of the
electromagnetic energy it carries. Radiation of the particles leads to damping of
the wave. The efficiency of this
process depends on the physical conditions imposed by the external medium. Two regimes can be
distinguished: a high density one, where the EM wave cannot be launched at all and
the current sheets in the striped wind are first compressed by an MHD shock and
subsequently dissipate by reconnection, and a low density one, where the wind can
first convert into an electromagnetic wave in the shock precursor, which then damps
and merges into the surroundings. Shocks in binary systems can transit from one
regime to another according to binary phase. We discuss possible observational
implications for these objects.
\end{abstract}

\maketitle


\section{Introduction: the $\sigma$-problem}

A wind, outflowing from a pulsar magnetosphere, is a mixture of plasma, produced close to the star, and the strong electromagnetic fields that are anchored in its surface and twisted beyond the light cylinder $r_{\rm L}=c/\omega$ ($\omega$ is the angular velocity of the star). This wind  carries the entire pulsar spin down power $L$, expressed via the dimensionless parameter $a_{\rm L}=(e^2L/m^2c^5)^{1/2}$, and the mass loading of the wind is given by $\mu=a_{\rm L}/4\kappa$, where $\kappa$ is the multiplicity of pulsar cascades. The energetics of the outflow is completely dominated by the Poynting flux, so that the magnetization parameter, the ratio of the energy flux carried by the fields to that carried by the particles, is large $\sigma\gg 1$. 
Within the ideal MHD description of the wind, a large initial value of $\sigma$ remains so along the entire flow, up to the point where the wind impacts on the surroundings and terminates in the standing shock. However, the observed emission, attributed to the synchrotron processes downstream of this shock, suggests that most of the energy is deposited in the particles. 
The mechanism that transfers the energy stored in the fields to the plasma is not clear, and called the \lq\lq $\sigma$-problem\rq\rq. A solution must be envisaged beyond the domain of validity of the ideal MHD approach. 
In particular, 
as the wind propagates radially outwards, the plasma density decreases, and beyond the critical radius $r_{\rm c}=a_{\rm L}r_{\rm L}/\mu$, it is low enough to enable the propagation of strong electromagnetic (EM) waves. 
Such an underdense medium, in contrast to the MHD plasma, cannot screen out the wave displacement currents, and when they take over, mode conversion between the MHD wind and the EM mode can occur. This transition can be regarded as a consequence of the boundary conditions, 
and the new EM wave mimics a precursor to the shock, launched to adjust the flow to the surroundings. This mode is able to efficiently 
transfer the EM energy to the plasma, providing a solution to the $\sigma$-problem.

\section{Electromagnetic shock precursors}

\paragraph{Properties and radial propagation of large-amplitude waves}

Propagation of strong EM waves is described within the two-fluid model of a plasma that is coupled to the EM fields via equations of motion and Maxwell equations. Under pulsar conditions the plasma is positronic, which implies that the electric vector of EM waves is purely transverse. The $e^{\pm}$ fluids move with equal momentum $\ppar=\gamma v_{\parallel}/c$ in the propagation
direction, but have oppositely directed, equal amplitude oscillations in transverse momentum $\pperp$. 
In the lab. frame these waves have superluminal phase velocity $\beta>1$, but subluminal group speed $\betaw=1/\beta<1$ that, in general, does
not coincide with the parallel component of the fluid 3-velocity $v_{\parallel}/c$ (i.e., the particles stream through
the wave). 

Using perturbation analysis in the small parameter $\epsilon=r_{\rm L}/r\ll 1$ (short wavelength approximation; $r$ is the distance at which the wave is probed), 
we expand all the quantities in order to obtain a set of equations that, to the lowest order, describe the plane wave solution, dependent only on the wave phase. To first order they govern the slow radial evolution of
the phase-averaged, plane-wave quantities. The radial propagation is given by (1) the continuity equation,
(2) energy conservation, and (3) the evolution of the radial momentum flux. In contrast to the
MHD wave, the radial momentum flux is not conserved in spherically expanding EM modes, because 
the flow exerts a pressure in the lateral direction, and the energy division between the parallel and these perpendicular degrees of freedom changes.  
In addition to the conserved particle and energy fluxes, it can be
shown \cite{mocholphd} that the phase-averaged Lorentz factor of the particles, measured in the laboratory
frame, $\av{\gamma_{\rm lab}}$, is an integral of motion for both circularly and linearly polarized modes.

\begin{figure}[!h]
\resizebox{0.5\columnwidth}{!}{\input{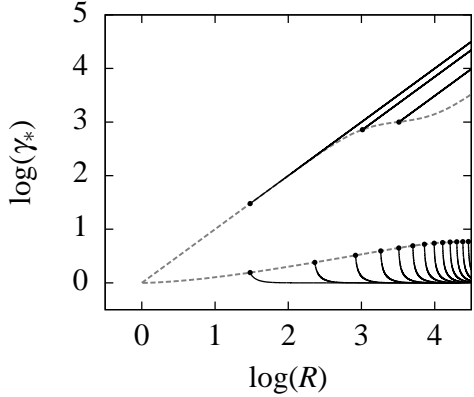}  \label{figconv}}
  \caption{Dashed: The Lorentz factor of a circularly polarized EM wave,  
corresponding to an MHD wind with 
$\mu=10^4$, $\sigma=100$, $a_{\rm L}=3.4\times10^{10}$; against conversion radius.
Solid: the radial evolution of an EM wave launched at a point on the 
dashed curve (marked by a dot). Radius is normalized $R=r\mu\omega/(ca_{\rm L})$.}
\end{figure}

The EM mode at the conversion radius is determined from 
the jump conditions between the MHD and the EM wave, which 
carry the same particle, energy and radial momentum
fluxes \cite{2010PPCF...52l4029K, 2012ApJ...745..108A}. 
Fig.~\ref{figconv} 
shows the Lorentz factor of a strong EM wave
$\gammaw$
, obtained from the jump
conditions (dashed curves), and its radial evolution (solid curves) for
different launching radii. There are two solutions of the jump
conditions that describe two possible EM modes: a free-escape mode
(higher branch) and a confined mode (lower branch). 
Only the latter 
decelerates at large distances, and, since the wind 
is terminated in the (roughly) standing shock, in the following we concentrate on this mode.
%
%
%

The unique determination of  
a wave at launch requires the knowledge of four quantities: 
the conversion radius $R_0$, initial group speed $\betawb$, and initial 
particle momenta $\pparb$, $\pperpb$. The jump conditions define three of 
them, leading to the dotted curves in
Fig.~\ref{figconv}. However, the evolution equations of the EM mode imply that its ram pressure tends to a constant value at large radius, which, in fact, should be equal to the external pressure $\pext$. Since only the wave launched at the correct radius has the correct asymptotic pressure, the fourth quantity, $R_0$, is determined by the matching procedure.  
Given that $\av{\gamma_{\rm lab}}$ is conserved, one can  
connect the
asymptotic wave pressure with the initial wave parameters. 
Therefore, a unique, 
stationary precursor-solution is determined by the MHD wave parameters 
$\sigma$, $\mu$, $a_L$, 
and 
also $\pext$, unless $p_{\rm ext}$ is larger than 
the ram pressure at $r=r_{\rm c}$, $p(r_{\rm c})\equiv p_{\rm c}$, in which case no such
precursor can be formed. 

\paragraph{Radiation damping of electromagnetic waves}

Self-consistent EM waves are prone to damping by nonlinear inverse
Compton (NIC) scattering (i.e., radiation reaction of the particles accelerated in the wave fields)
\cite{1978A&A....65..401A}, as well as inverse 
Compton scattering of ambient low energy photons.  Even though the flow 
energy is dissipated to the photon field, the 
particles themselves gain more
energy from the fields than they lose in radiation, and both the
wave and the particle streaming slow down due to the interaction. Thus, this effect resembles that
caused by radial expansion. When, by either of these mechanisms, the
streaming of the particles through the wave vanishes, strong waves become very unstable to small
density perturbations in the direction of motion
\cite{1973PhFl...16.1480M,1978JPlPh..20..313L,1978JPlPh..20..479R}.
Parametric instabilities then set in and destroy the wave in only one wave period. This
point of vanishing streaming is the location of the termination shock, beyond which the flow
energy is thermalized. 

\section{Implications for the emission in binaries}

For specified values of the magnetospheric and wind parameters, the shock structure is determined only by the pressure conditions in the external medium. We identify two regimes: 
\begin{description}
\item[one with high external pressure $\pext>p_{\rm c}$] -- superluminal mode cannot be launched, because the conversion radius, required by the external conditions, falls below the critical radius; the current sheets in the pulsar wind are first compressed by an MHD shock and
subsequently dissipate by reconnection \cite{2003MNRAS.345..153L},
\item[one with lower external pressure $\pext<p_{\rm c}$] -- the EM precursor can be launched; in this case two possibilities arise:

(1) $\pext\leq p_{\rm c}$ -- radiation damping of a wave is efficient and determines the shock location at the point of vanishing streaming,

(2) $\pext\ll p_{\rm c}$ -- radiation damping ineffective; the particle streaming decreases due to, e.g., radial expansion, until at large distances it drops sufficiently, the intrinsic instabilities set in and disrupt the wave, forming a shock.
\end{description}

In binaries, the high-pressure environment is provided by a wind of a companion star. If 
in the periastron $\pext<p_{\rm c}$, so that the precursor can be launched, it has a large amplitude, and therefore the NIC process very quickly decreases the wave streaming. 
In this case the shock is formed without significant emission from the precursor. Most of the flow energy can be deposited in the shock-accelerated particles and one can expect enhanced emission from the shock itself. 
Another possibility is that at periastron $\pext>p_{\rm c}$, and the precursor can be launched only in a more distant part of the orbit. 
At larger orbital separation, the damping is less abrupt and the emission from the precursor may constitute a significant fraction of the radiative output of the system. 

In the binary PSR B1259-63 the pulsar is on a very elongated orbit around a companion star and the external conditions for the pulsar wind change with the orbital phase. $\pext$ can be determined from the standard model of Be-star winds, e.g., \cite{2006A&A...456..801D}. 
The precursor can exist if the value of $\mu$ is high enough to support the EM wave solutions required by $\pext$.  
For this system, if $\mu\sim10^5$ the precursor can be launched in every point of the orbit. NIC emission in the periastron would be in the X-ray band, and in the optical band in the apastron -- possibly hidden beneath the luminous stellar emission. However, if $\mu\sim10^4$, the shock regime may be switched, with the MHD structure close to the periastron, and an EM, non-MHD precursor further out. If the regime switching occurs 30 days after the periastron passage, when the binary members are $3\times10^{13}$ cm apart, the critical radius equivalent to this distance would imply $\mu=2\times10^4$. This is equivalent to the multiplicity $\kappa\sim10^5$, about an order of magnitude smaller than the one inferred for the Crab pulsar. The emission from the precursor would be in the optical band, and the IC emission associated with the particles upscattering the stellar photons, in a few hundred MeV -- consistent with the high energy flare observed by Fermi-LAT \cite{2011ApJ...736L..10T}.

\bibliographystyle{aipproc}   

\bibliography{proc}

\begin{thebibliography}{10}
\expandafter\ifx\csname natexlab\endcsname\relax\def\natexlab#1{#1}\fi
\providecommand{\enquote}[1]{``#1''}
\expandafter\ifx\csname url\endcsname\relax
  \def\url#1{\texttt{#1}}\fi
\expandafter\ifx\csname urlprefix\endcsname\relax\def\urlprefix{URL }\fi
\providecommand{\eprint}[2][]{\url{#2}}

\bibitem[{Mochol}(2012)]{mocholphd}
I.~{Mochol}, \emph{Nonlinear waves in Poynting-flux dominated outflows}, Ph.D.
  thesis, University of Heidelberg (2012).

\bibitem[{Kirk}(2010)]{2010PPCF...52l4029K}
J.~G. {Kirk}, \emph{Plasma Physics and Controlled Fusion} \textbf{52}, 124029
  (2010), \eprint{1008.0536}.

\bibitem[{Arka} and {Kirk}(2012)]{2012ApJ...745..108A}
I.~{Arka}, and J.~G. {Kirk}, \emph{\apj} \textbf{745}, 108 (2012),
  \eprint{1109.2756}.

\bibitem[{Asseo} et~al.(1978)]{1978A&A....65..401A}
E.~{Asseo}, C.~F. {Kennel}, and R.~{Pellat}, \emph{\aap} \textbf{65}, 401--408
  (1978).

\bibitem[{Max}(1973)]{1973PhFl...16.1480M}
C.~E. {Max}, \emph{Physics of Fluids} \textbf{16}, 1480--1489 (1973).

\bibitem[{Lee} and {Lerche}(1978)]{1978JPlPh..20..313L}
M.~A. {Lee}, and I.~{Lerche}, \emph{Journal of Plasma Physics} \textbf{20},
  313--328 (1978).

\bibitem[{Romeiras}(1978)]{1978JPlPh..20..479R}
F.~J. {Romeiras}, \emph{Journal of Plasma Physics} \textbf{20}, 479--501
  (1978).

\bibitem[{Lyubarsky}(2003)]{2003MNRAS.345..153L}
Y.~E. {Lyubarsky}, \emph{\mnras} \textbf{345}, 153--160 (2003),
  \eprint{arXiv:astro-ph/0306435}.

\bibitem[{Dubus}(2006)]{2006A&A...456..801D}
G.~{Dubus}, \emph{\aap} \textbf{456}, 801--817 (2006),
  \eprint{arXiv:astro-ph/0605287}.

\bibitem[{Tam et al.}(2011)]{2011ApJ...736L..10T}
P.~H.~T. {Tam et al.}, \emph{\apjl} \textbf{736}, L10 (2011),
  \eprint{1103.3129}.

\end{thebibliography}

\IfFileExists{\jobname.bbl}{}
 {\typeout{}
  \typeout{******************************************}
  \typeout{** Please run "bibtex \jobname" to optain}
  \typeout{** the bibliography and then re-run LaTeX}
  \typeout{** twice to fix the references!}
  \typeout{******************************************}
  \typeout{}
 }

\end{document}